\documentstyle[ppeuc,epsfig]{article}
\newcommand{\Omo}{\Omega_{\rm o}}
\begin{document}

\title{Holes in the Microwave Sky}
\author{\addlink{J.G. Bartlett}{bartlett@astro.u-strasbg.fr}\\
	\addlink{A. Blanchard}{blanchard@astro.u-strasbg.fr}\\
	\addlink{D. Barbosa}{barbosa@astro.u-strasbg.fr}}
\institute{\addlink{Observatoire de Strasbourg}{http://astro.u-strasbg.fr/
	Obs.html}, 11, rue de l'Universit\'e\\
           67000 Strasbourg, FRANCE}
\author{\addlink{J. Oukbir}{jamila@dsri.dk}}
\institute{Danish Space Research Institute, DSKI Juliane Maries Vej 30\\
	DK-2700, Copenhagen 0, DENMARK}

\begin{abstract}
We discuss the implications of two possible, recent Sunyaev-Zel'dovich (SZ)
detections for which no optical or X--ray counterparts have been 
found.  This suggests that the objects reside at high redshift,
which is difficult to reconcile with a critical cosmology.  We
develop this argument and find that an open model with $\Omo=0.2$
remains consistent with what we currently know about the
two objects.  The reasoning also demonstrates the utility of 
SZ cluster searches.
\end{abstract}

\section{Introduction}

	By holes in the microwave sky, we hope to invoke an image
of the Sunyaev--Zel'dovich (SZ) effect at low frequences: a region
of lower than average sky brightness, or a hole in the cosmic
microwave background (CMB).  The image presented by B. Partridge
at these proceedings is an excellent example and, along with
a similar detection by the Ryle Telescope, is the motivation
for this presentation; because if the two radio
decrements are indeed due to SZ effect, this can have
powerful implications for the value of $\Omo$.

	In what has yielded the deepest radio map to date, 
the VLA discovered a radio decrement -- characteristic of
the SZ effect below 1.4 mm -- during an observation of one 
of the HST Medium Deep Survey fields (Richards et al. \cite{vla}).
The object is just resolved, extending over an area of 
about $30\times 60\; {\rm arcsec}^2$.  The other object 
(Jones et al. \cite{ryle1}) was found by the RYLE Telescope (RT)
during an ongoing program of double quasar observations
(Saunders \cite{ryle2}).
They find a radio decrement covering an area of about 
$120\times 180\; {\rm arcsecs}^2$.
In both cases, subsequent follow--up in the optical and 
in the X--ray band has not reveiled the supposed clusters
(Richards et al. \cite{vla}; Jones et al \cite{ryle1}; 
Saunders et al. \cite{ryle3}).
Definite confirmation of the SZ nature of the two decrements
will thus come from efforts to measure the effect at different
frequencies, to see if the spectra are consistent with
the SZ effect.  If they are indeed clusters, then the
lack of optical or X--ray counterparts may be interpretated
as evidence that they lie at large redshift.  It is
in this way that we may obtain very strong constraints on
$\Omo$:  The number of massive, high--redshift clusters depends
sensitively on $\Omo$, so much so that the observation
of even a small number of such clusters can eliminate the
critical model (Oukbir \& Blanchard \cite{ob92}; 
Barbosa et al. \cite{SZcounts}; Eke et al. \cite{eke};  
Oukbir \& Blanchard \cite{ob}).  

\section{The Mass Function and $\Omo$}

	The easiest way to understand this dependence is by
considering the mass function, the number density of
collapsed objects as a function of mass and redshift (e.g., Bartlett 
\cite{casa}).  
In `standard' models, based on the growth of initially small density
perturbations with gaussian statistics, this takes the 
general form of a power law times a gaussian (Press \& Schechter \cite{PS}):
\begin{equation}
\label{eq:PSfunc}
\frac{dn}{d{\rm ln} M} = \sqrt{\frac{2}{\pi}}\frac{<\rho>}{M}\nu(M,z)
	\left(-\frac{d{\rm ln} \sigma(M,z)}{d{\rm ln}M} \right)
	e^{-\nu^2/2}.
\end{equation}
In this equation, $<\rho>$ is the mean, comoving density
of the Universe and $\sigma(M,z)$ is the power spectrum as a 
function of the mass scale, $M$.  The quantity
\begin{equation}
\nu(M,z) = \frac{\delta_c(z)}{\sigma_{\rm o}(M)} \frac{D_{\rm g}(z=0)}
	{D_{\rm g}(z)}
\end{equation}
is a function of the linear growth factor, $D_{\rm g}(z;\Omo,\Lambda)$,
which depends on $\Omo$ and $\Lambda$, of the critical density
needed for collapse, $\delta_c$, which has only a weak dependence on
$\Omo$ and $\Lambda$, and of $\sigma_{\rm o}(M)$, 
the present--day power spectrum,
a function of mass only.  The appearance of $D_{\rm g}$ in
the exponential of the mass function indicates that the 
$\Omo$ dependence can be quite strong; hence, the comment
that even a small number of clusters at large $z$ can 
severely constrain the density parameter.  The key 
point is that {\em the shape of the redshift distribution of 
clusters of a given mass is only determined by the cosmological 
parameters} (the power spectrum cannot be changed to alter this
fact) (Oukbir \& Blanchard \cite{ob}).  

	The problem is that we do not measure mass directly;
we need some other, more readily observable quantity
which correlates well with cluster mass.  Because we 
believe that the hot cluster gas is heated by infall during
cluster formation, we expect that the X--ray temperature
should represent the depth of the cluster potential well and,
therefore, its mass.  This has in fact been well established
by various hydrodynamical simulations (in `standard'
scenarios) (Evrard et al. \cite{Tsim}), which also provide the exact 
form of the temperature--mass relation.  The X--ray luminosity,
on the other hand, is a more complicated animal, depending
not only on the temperature of the gas, but also on its
abundance and spatial distribution.  As we discuss below,
observing clusters via the Sunyaev--Zel'dovich effect 
avoids these problems associated with the X--ray flux,
while preserving the simplicity of a straightforward flux measurement 
(plus other advantages).  This is important because X--ray 
spectra demand time--consuming, space--based observations. 

\begin{table}
\begin{center}
{\bf Table}\\
Model Parameters - normalized to the local X--ray temperature function$^a$ \\
\begin{tabular}{*4{|c}|}
\hline
$\Omo$ & $h$ & $\sigma_8$ & $n$   \\ 
\hline  
   0.2 & 0.5 & 1.37       & -1.10 \\
   1.0 & 0.5 & 0.61       & -1.85 \\
\hline
\end{tabular}

$a$ - Henry \& Arnaud (\cite{ha})
\end{center}
\end{table}

	For our discussion here, we take a phenomenological 
point--of--view and adopt a power--law approximation to the power 
spectrum:  $\sigma_{\rm o} = (1/b)(M/M_8)^{-\alpha}$, where $b$ is the
bias parameter and $M_8$ is the mass contained in a sphere
of $8\; h^{-1}\; {\rm  Mpc}$.  We will focus on the comparison of two
extreme models, a critical model and an open model with
$\Omega=0.2$ ($h=H_o/100\; {\rm km/s/Mpc} = 1/2$ in both cases).  
The parameters
$b$ and $\alpha$ for each model are constrained by fitting
to the local X--ray temperature function of galaxy clusters
(Henry \& Arnaud \cite{ha}), the results of which are given in the
Table (Oukbir et al. \cite{obb}; Oukbir \&
Blanchard \cite{ob}).

\section{The Sunyaev-Zel'dovich Effect}

\begin{figure}
\label{fig:zdist}
\begin{center}
\psfig{figure=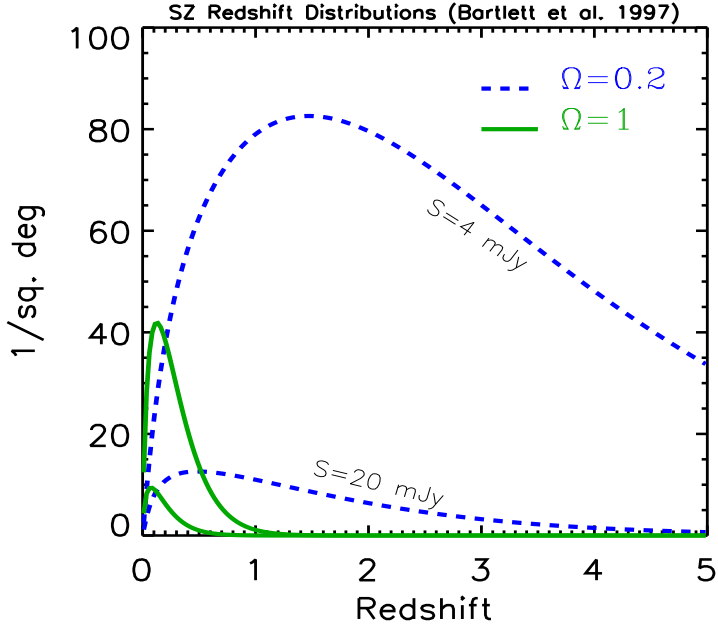,width=10cm,height=9cm,clip=}
\caption{Redshift distribution for clusters of given SZ flux
density in the two cosmological models -- critical (solid
lines) and open with $\Omega=0.2$ (dashed lines);  
see the Table for the model parameters.   
The curves are drawn for $S_{sz}=4\; {\rm mJy}$ and 
$S_{sz}=20\; {\rm mJy}$, corresponding to the VLA and RT objects,
respectively, when translated to $\lambda=0.75\; {\rm mm}$.
For clarity, the two curves for the critical model are not 
labeled explicitly.  We adopt $h=1/2$.}
\end{center}
\end{figure}

	The Sunyaev-Zel'dovich effect offers unique
advantages for finding high redshift clusters and quantifying
their abundance.  The surface brightness of a cluster 
relative to the unperturbed CMB is expressed as a product
of a spectral function, $j_\nu$, and the {\em Compton
$y$--parameter}, which is an integral of the electron pressure
along the line--of--sight: $y\propto \int dl\; n_e T$.
Integrating the surface brightness over solid angle yields the 
following functional form for the total flux density of a cluster
with angular size $\theta$:
\begin{eqnarray}
\label{eq:SZflux}
\nonumber
S_{sz} \propto \theta^2 n_e T l \propto M_{gas} T D_a^{-2}(z)\\
\propto f_{gas} M_{tot}^{5/3} (1+z) D_a^{-2}(z),
\end{eqnarray}
where $M_{tot}$ is the total, virial mass of the cluster,
$f_{gas}$ is the hot gas mass fraction of clusters and 
$D_a(z)$ is the angular--size distance.  In the last line, 
we have used the fact that there exists a tight relation
between X--ray temperature and virial mass: 
$T \propto M^{2/3} (1+z)$ (Evrard et al. 1996).  
Let's compare this with the corresponding expression for 
the X--ray flux of a cluster:
\begin{eqnarray}
\label{eq:fx}
\nonumber
f_x = \theta^2 n_e^2 T^{1/2} l (1+z)^{-4} 
	\propto n_e M_{gas} T^{1/2} D_a^{-2}(z) (1+z)^{-4} \\
\propto n_e f_{gas} M_{tot} T^{1/2} D_l^{-2}(z),
\end{eqnarray}
with $D_l$ denoting the luminosity distance.  By comparing these
two expressions we see that, in contrast to the SZ flux density,
the X--ray flux suffers cosmological surface brightness dimming, 
represented by the extra factors of $(1+z)$ in the denominator
of Eq. (\ref{eq:fx}) which convert the angular--size distance to 
the luminosity distance.  Besides this well--known 
difference, which tells us that the SZ effect is the more
efficient way to find high--redshift clusters, we note
that the X--ray emission depends on the gas density in 
addition to the hot gas mass fraction and temperature.  
This is unfortunate, because it means that the X--ray
flux from a cluster depends on the core radius and profile of 
the intracluster medium (ICM) -- two quantities which are
poorly, if at all, understood from the theoretical point
of view.  The SZ flux density presents the important advantage
that it depends {\em only on the total gas mass and the 
temperature}, and not on the ICM's distribution.  It is
also true that the temperature which appears in the expression
for the SZ flux density is a simpler quantity than the X--ray
measured temperature: it is the {\em mean, particle--weighted}
energy of the gas particles instead of, as in the case of
X--rays, the emission--weighted gas temperature.  This 
SZ temperature is a quantity which should be all the more 
closely related to the virial mass of a cluster than even the
X--ray temperature, and less affected by any temperature
structure in the cluster.

\begin{figure}
\label{fig:SZcounts}
\begin{center}
\psfig{figure=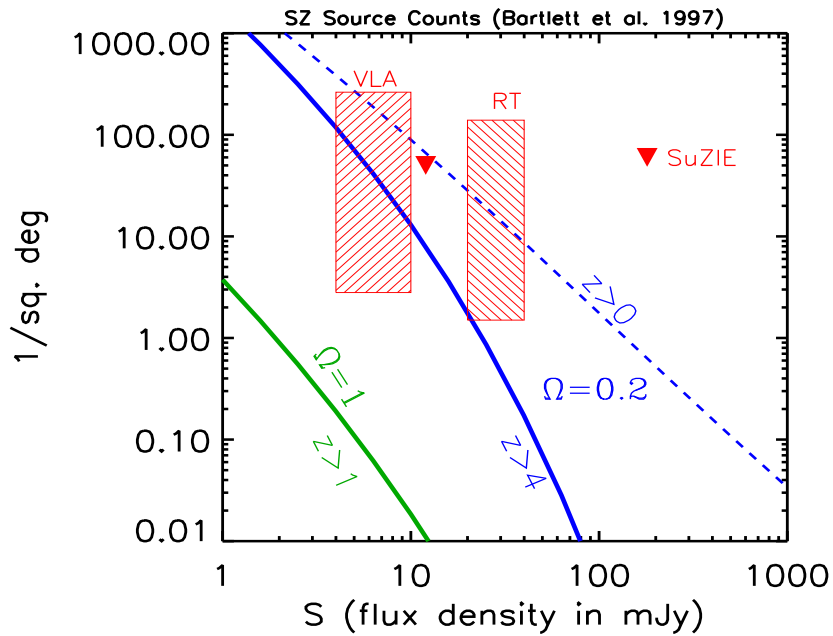,width=10cm,height=9cm}
\caption{SZ source counts with observational constraints,
as a function of SZ flux density expressed at 
$\lambda=0.75\; {\rm mm}$.  The two hatched boxes show the 
95\% {\em one--sided} confidence limits from the VLA and 
the RT; due to the uncertain redshift of the clusters, 
there is a range of possible {\em total} SZ flux density,
which has for a minimum the value observed in each beam
and a maximum chosen here to correspond to $z>1$.  From 
the SuZIE blank fields, one can deduce the 95\% upper limit
shown as the triangle pointing downwards 
(Church et al. \cite{suzie}).  We also plot the predictions
of our fiducial open model ($\Omega=0.2$) for all clusters 
(dashed line) and for those clusters with $z>4$.  The critical model
has great difficulty explaining the observed objects
even with a lower redshift cutoff of only $z>1$; the actual
limit from the X--ray data is stronger, but this would
fall well off to the lower left of the plot.  We assume
$h=1/2$.}
\end{center}
\end{figure}

	Now the game is clear: with Eq. (\ref{eq:SZflux}) we
may convert the mass function into a distribution of clusters
in SZ flux density and redshift (the quantitative
relation for $S_{sz}$ can be found in Barbosa et al. \cite{SZcounts}).  
The redshift distribution of clusters and the total source
counts are then easily calculable (Korolyov et al. \cite{koro};
Markevitch et al. \cite{mark}; Bartlett \& Silk \cite{bs94}; 
Barbosa et al. \cite{SZcounts}; Eke et al. \cite{eke};
Colafrancesco et al. \cite{cola}).  In Figure 1, 
we show the redshift distribution for clusters of two given SZ 
flux densities and for two representative cosmologies -- a 
critical model and a model with $\Omega=0.2$ and $\Lambda=0$.
For this calculation, we have used a constant gas mass fraction
$f_{gas}=0.06\;h^{-3/2}\;$ (Evrard \cite{gasfrac}).
The two chosen fluxes are our estimates of the flux density
of the VLA and RT objects, when translated to a wavelength
of $\lambda=0.75\; {\rm mm}$ by using the SZ spectral function,
$j_\nu$;  this is our fiducial working frequency and 
corresponds to the peak of the SZ distortion.  

	The key aspect of this figure is that, at a given flux 
density, there is an enormous difference between the number of 
high--redshift clusters in a critical and open universe.  
It is for this reason that even the detection of only two SZ
decrements warrants the present discussion, because they
appear to be at large redshift.  Let us now quantify this
by comparing the predicted number counts of clusters 
with redshifts greater than some minimum value with the
counts implied by the detection of these two objects.
This is done in Figure 2.  The observed
counts have been estimated using Poisson statistics
and the amount of sky coverage in each case --
$\sim 0.018\; {\rm deg}^{2}$ for the VLA (two fields) and
$\sim 0.034\; {\rm deg}^{2}$ for the RT (three fields). 
These constraints are given as two boxes because
there is actually a range of possible {\em total} 
SZ flux density from each object, due to the unknown
redshifts: at low redshift, the objects would be 
resolved and their flux density has to be corrected upwards.
The minimum flux density is clearly the value observed,
while for the maximum, we give the values for $z\sim 1$
assuming an isothermal $\beta-{\rm model}$. 
The limits on the counts (i.e., in the vertical direction)
are generous in that they represent
the 95\% {\em one--sided} Poisson confidence limits.
We also show an upper limit on the counts obtained by
the SuZIE instrument (Church et al. \cite{suzie}), which
found no objects in a survey area of $\sim 0.06\; {\rm deg}^{2}$
down to the limiting flux shown; the symbol represents
the resulting 95\% confidence upper limit. 
Predictions for the number of clusters on the sky
for the two cosmological models and with varying minimum 
redshifts are shown by the labeled curves.

\section{Conclusions}

	The basic result from Figure 2 is clear:
if the two radio decrements are indeed due to the thermal
SZ effect in two clusters, then the critical model is 
in very serious trouble.  On the other hand, an open model
is capable of explaining the objects.  While our modeling has
been rather simple in that we have used power--law power
spectra and assumed that the cluster gas fraction is
constant over mass and epoch (see, e.g., Colafrancesco et al.
\cite{cola} for more detailed treatment of cluster
evolution), in the present circumstance 
any reasonable evolution in the gas mass fraction would
lead to a decrease in the SZ flux density of objects with 
small mass and/or at large redshift; hence, it would only make
things {\em more difficult} for the critical model.    

	What about other possible caveats?  Besides the
fact that we still await definite confirmation of the true
nature of the radio decrements (e.g., detections at other
frequencies), the most important thing to be wary of is the possible
bias associated with the fact that the RT was pointed at
a known double quasar, not an a priori blank field.  This
seems less likely to be true of the VLA detection, in as much
as there was no previously known quasar pair in that field
(although one was subsequently found).  In any case, 
we should have a better idea in the near future from 
other experiments as the amount of sky covered by SZ
searches appears to be rapidly increasing.  The present
discussion brings to light the importance of such
SZ searches (we note in particular that a $\sim 1$ square--degree 
search with the BIMA telescope is now feasible [Holzapfel, private communication]).  A satellite mission covering the full sky,
such as the Planck Surveyor, will be the culmination of such
efforts.  

	Finally, we remark that the possible existence
of clusters at redshifts much greater than unity should
not be seen as exotic; quite the contrary, in open models,
they are expected.  If they are indeed out there, they
would not have been detectable up until now by either
optical or X--ray observations.  One would imagine that
they would first be seen by SZ searches, and these 
are just now beginning to provide some very interesting
and tantalizing hints.

\section*{Acknowledgements}

	We thank the organizers for a very interesting and enjoyable
meeting.  We would also like to thank B.~Partridge for some helpful
and pleasant discussions.

\end{document}